\documentclass[11pt]{article}
\usepackage{vietnam,epsfig}

\def\be{\begin{equation}}
\def\ee{\end{equation}}
\def\bea{\begin{eqnarray}}
\def\eea{\end{eqnarray}}

\begin{document}
\vspace*{4cm}
\title{Propagation and Chemical Composition of Ultra High Energy Cosmic Rays}

\author{Roberto Aloisio}

\address{INFN - Labratori Nazionali del Gran Sasso, I-67010 Assergi (AQ),
 Italy}

\maketitle\abstracts{
Extragalactic cosmic ray protons with an injection spectrum of the type $E^{-2.7}$
show a spectrum on earth with a dip due to the Bethe-Heitler pair production against the photons of the cosmic microwave background. The dip is produced in the energy region 
$10^{18} - 4\times 10^{19}$ eV with position and shape that reproduce with high accuracy
the spectrum observed experimentally. This interpretation of the observed data predicts the existence of an energy scale that signals a possible transition from galactic to extragalactic cosmic rays. In fact, at energies lower than a characteristic value $E_c\approx 1\times 10^{18}$ eV, determined by the equality between the rate of energy losses due to pair production and adiabatic losses, the spectrum of cosmic rays flattens in all cases of interest. In this model, the transition from galactic to extragalactic cosmic rays occurs at some energy below $E_c$, corresponding to the position of the so-called second knee. Another viable explanation of the observed data is based on a completely different approach assuming a mixed composition with protons and nuclei at energies 
$E\ge 10^{19}$ eV. This scenario implies a transition from galactic to extragalactic cosmic rays 
at the ankle energies ($\sim 10^{19}$ eV). In the present paper we will review the main features of the dip model comparing it with the model of transition at the ankle. 
}

\section{Introduction}
\label{sec:intro}

A clear understanding of the provenance of Ultra High Energy Cosmic Rays (UHECR) with 
$E \ge 10^{18}$ eV, namely their galactic or extragalactic origin, will be of paramount importance in unveiling the nature of the sources of these particles. Assuming that extragalactic cosmic rays are 
dominated by protons, the propagation in the intergalactic medium induces 
two features in the spectrum that can be observed on Earth: 1) the GZK feature 
\cite{GZK}, a suppression of the flux at energies in excess of $\sim 10^{20}$ 
eV, due to the photopion production of cosmic rays off the Cosmic Microwave Background (CMB) photons; 2) A dip, \cite{BGG,dip} generated due to pair production, $p+\gamma_{\rm CMB} \to p+e^++e^-$, where the target is provided by the CMB photons.

The detection of these features can be interpreted as a test of the
extragalactic origin of UHECRs and of the fact that they are mainly protons.
Since the detection of the GZK feature requires very large statistics 
of events, at present, the spectral feature that can be detected 
more easily is the dip. As we discuss here (but see also \cite{BGG,dip}), 
the dip is a quite robust prediction of the calculation and in fact 
it might have already been observed by the AGASA, Fly's Eye, HiRes and Yakutsk 
experiments (see \cite{dataUHECR} for the data and \cite{NW} for a review). 

However, the detection of the dip as a feature of the propagation of 
cosmic rays on cosmological scales would also imply that the transition 
from galactic to extragalactic cosmic rays should take place at energies 
below $10^{18}$ eV and not at larger energies as has been postulated so far. 
In particular there are two competitive scenarios: the dip scenario and the mixed composition scenario \cite{Allard}. The mixed composition model assumes a transition between galactic and extra-galactic Cosmic Rays (CR) at energies around $3-5\times 10^{18}$ eV, with galactic nuclei at energies $E<3\times 10^{18}$ eV and extra-galactic mixed protons and nuclei at larger energies. 

The mixed composition scenario resembles more like the traditional explanation of the transition from galactic to extragalactic cosmic rays. The traditional approach, introduced in the seventies, invokes the intersection between a steep ($E^{-3.1}$) galactic spectrum and a flat 
($E^{-\alpha}$ with $\alpha=2-2.3$) extragalactic spectrum, at the so-called ankle, located at an energy $E_a \approx 10^{19}$ eV and identified as a flattening of the spectrum in the data of AGASA, HiRes and Yakutsk detectors.

In the dip model the predicted spectrum of cosmic rays immediately below and above the dip location flattens: the high energy flattening, at $E_a \approx 1\times 10^{19}$ eV, coincides 
with the well known and well observed ankle. The low energy flattening, 
at $E_c \approx 10^{18}$ eV, obtained in both cases of rectilinear \cite{BGG}
and diffusive propagation \cite{AB,Lem}, defines the region where
the transition from galactic to extragalactic (protons) CR takes place. 

In the present paper we discuss (as demonstrated in  \cite{AB}) how the critical energy $E_c$ 
is connected with the energy scale $E_{\rm eq} = 2.3 \times 10^{18}$ eV, where 
the rates of pair production and adiabatic energy losses are equal.  
The visible transition from galactic to extragalactic cosmic rays occurs at 
$E_{\rm tr} < E_c$, and this energy coincides with the position of
the {\em second knee}  (Akeno - $6\times 10^{17}$~eV, Fly's Eye - 
$4\times 10^{17}$~eV, HiRes - $7\times 10^{17}$~eV and 
Yakutsk - $8\times 10^{17}$~eV). 

The energy region around $E_c \approx 10^{18}$ is also expected to 
correspond to a transition in the chemical composition, from a heavy 
galactic component to a proton-dominated extragalactic component. While HiRes, 
HiRes-MIA and Yakutsk \cite{mass} data support this prediction and 
Haverah-Park \cite{dataUHECR} data do not contradict it at 
$E \ge (1 - 2)\times 10^{18}$~eV, the Akeno and Fly's Eye 
\cite{mass} data favor a mixed composition, dominated 
by heavy nuclei (for a review see \cite{NW} and \cite{Watson04}).

In the preset paper we will review the main features of the dip model comparing 
it with the mixed composition scenario. The paper is organized as follows: 
in section \ref{sec:thedip} we discuss the physics behind the formation of the dip, 
and compare our predictions with the data of Akeno and AGASA experiments. In  section 
\ref{sec:transition} we address the more specific issue of the transition from 
galactic to extragalactic cosmic rays conclusions take place  in section 
\ref{sec:conclusions}. 

\section{The Dip}
\label{sec:thedip}

In this section we discuss the physical arguments that explain the formation 
of the dip in the spectrum. The UHECR spectrum $J_p(E)$ can be calculated 
imposing the conservation of the number of particles as
\be
J_p(E,t_0)dE= \frac{c}{4\pi}\int_{t_{min}}^{t_0} dt Q_{\rm gen}(E_g,t)dE_g,
\label{conserv}
\ee
where $n_p(E,t_0)$ is the space density of UHE protons at present, 
$t_0$, cosmological epoch, $Q_{\rm gen}(E_g,t)$ is the generation rate per
comoving volume at cosmological time t, and $E_g(E,t)$ is the generation
energy at time t for a proton with energy E at $t=t_0$. This energy  
is determined solving the evolution equation $dE/dt=b(E,t)$, where 
$b(E,t)$ are energy losses at epoch t. 

The spectrum (\ref{conserv}), calculated for an homogeneous distribution of 
sources, is called the {\em universal spectrum}. The important feature of 
the universal spectrum is its independence of the progation mode: it is the 
same for rectilinear propagation and propagation in arbitrary magnetic fields. 
This property of the universal spectrum is guaranteed by the propagation 
theorem \cite{AB}, according to which the spectra do not depend on the 
propagation mode if the distance between sources is less than any propagation 
length, e.g. energy attenuation or diffusion length. For homogeneous 
distribution of sources with vanishing distance between them the propagation 
theorem is obviously fulfilled.

\begin{figure}[ht]
\begin{center}
\begin{tabular}{ll}
\includegraphics[width=0.45\textwidth]{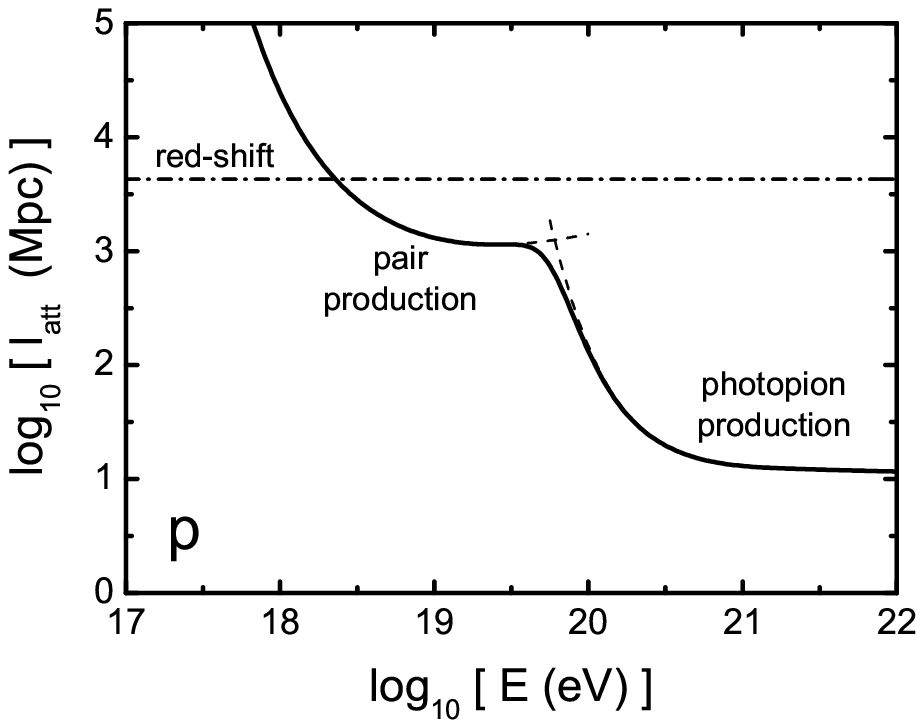}
&
\includegraphics[width=0.48\textwidth]{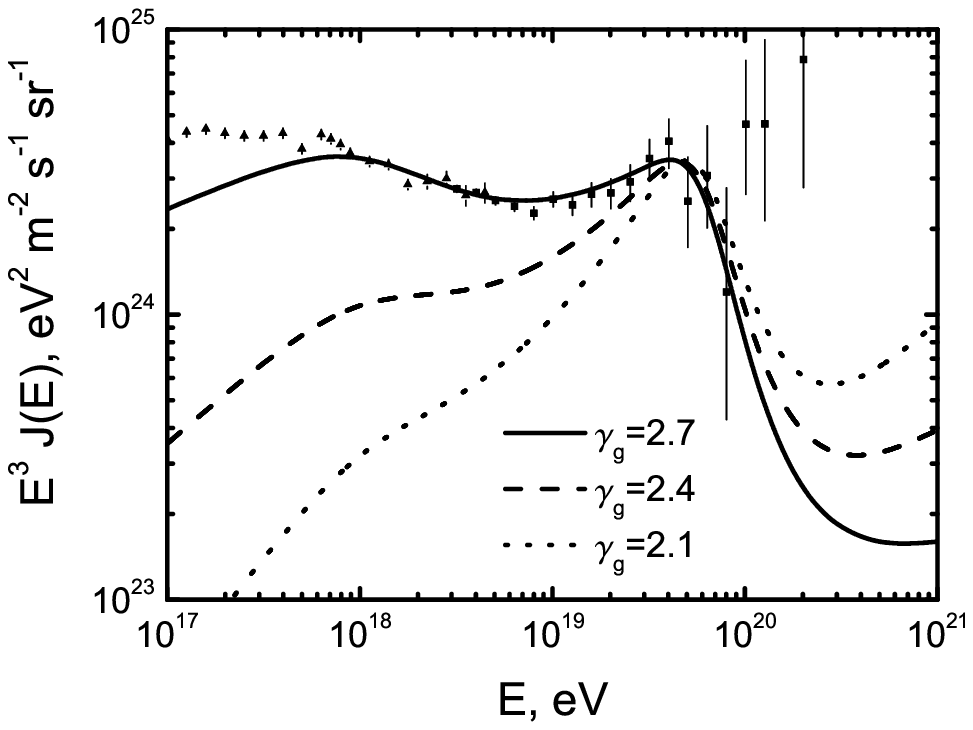}
\\
\end{tabular}
\caption{[Right Panel] Protons energy losses on the CMB background field.
[Left Panel] Spectra of protons for different choices of the injection spectrum, 
$\gamma_g=2.7$ continuos line, $\gamma_g=2.4$ dashed line and 
$\gamma_g=2.1$ dotted line. The experimental points are the Akeno-AGASA data.}
\label{fig:protons}
\end{center}
\end{figure}

In the right panel of figure \ref{fig:protons} we plot the energy losses $b(E,t_0)$ 
(at present) suffered by UHE protons, distinguishing the three different 
channels of adiabatic, pair-production and photopion production. In the 
left panel of figure \ref{fig:protons} we plot the UHECR spectrum obtained assuming a 
simple power law for the injection spectrum: $Q_{inj}\propto E_g^{-\gamma_g}$,
with three possible choices of the power law index $\gamma_g=2.1, 2.4, 2.7$.

As discussed in \cite{dip}, UHECR data can be described with an unprecedented accuracy assuming an injection spectrum with a power law index $\gamma_g=2.7$.
The excellent agreement of the predicted and observed position of the dip is clear in the case of AGASA (as shown by right panle of figure \ref{fig:protons}),  HiRes, Yakutsk and Fly's Eye data \cite{dip}. The case of Auger is, at the present time, rather ambiguous: the data do not contradict the presence of the dip, but the energy threshold is too high to show in a clear way the dip structure \cite{dip}. 

From the right panel of figure \ref{fig:protons} one can see the effect of energy losses 
on the proton spectrum. Using a steep injection spectrum, like the best fit value 
$\gamma_g=2.7$, the number of low energy particles is increased respect to the case of a less steep injections, as the two cases $\gamma_g=2.1$ and $\gamma_g=2.4$. Increasing the number of low energy particles the effect of the pair production process becomes more important producing the spectrum behavior shown by left panel of figure \ref{fig:protons}. 

Since the dip location in energy is defined rather precisely by the calculations, assuming it is related to the process of pair production, as discussed in \cite{dip}, it could serve as an exceptional
calibration tool for present and future experiments. 

The systematic errors in the energy determination of existing detectors is of the order of 20\% and sometimes in excess of this. In \cite{DBO} the authors used a detailed Monte Carlo simulations of the propagation of cosmic rays, with statistical errors taken into account and including the
possibility of a systematic error in the energy determination: the reached conclusion was that the alleged discrepancy between the AGASA and HiRes experiments could be explained in terms of a combination of statistical and systematic errors in the energy determination, and
statistically limited number of events at energies above $\sim 10^{20}$ eV. This conclusion was strenghtened in \cite{DBO1} where the same authors showed that the realizations of the simulated propagation of UHECR that are found to have 11 or more events at energy above $10^{20}$ eV
resemble very closely the AGASA data, although the {\it average} spectrum has a pronounced GZK feature. In the same paper, the authors also make an attempt to extract events at random directly from the AGASA data and calculate the probability of obtaining the HiRes spectrum by chance. In both cases the alleged discrepancy between the two experiments is found to be statistically not very significant.

\begin{figure}[ht]
\begin{center}
\begin{tabular}{ll}
\includegraphics[width=0.45\textwidth]{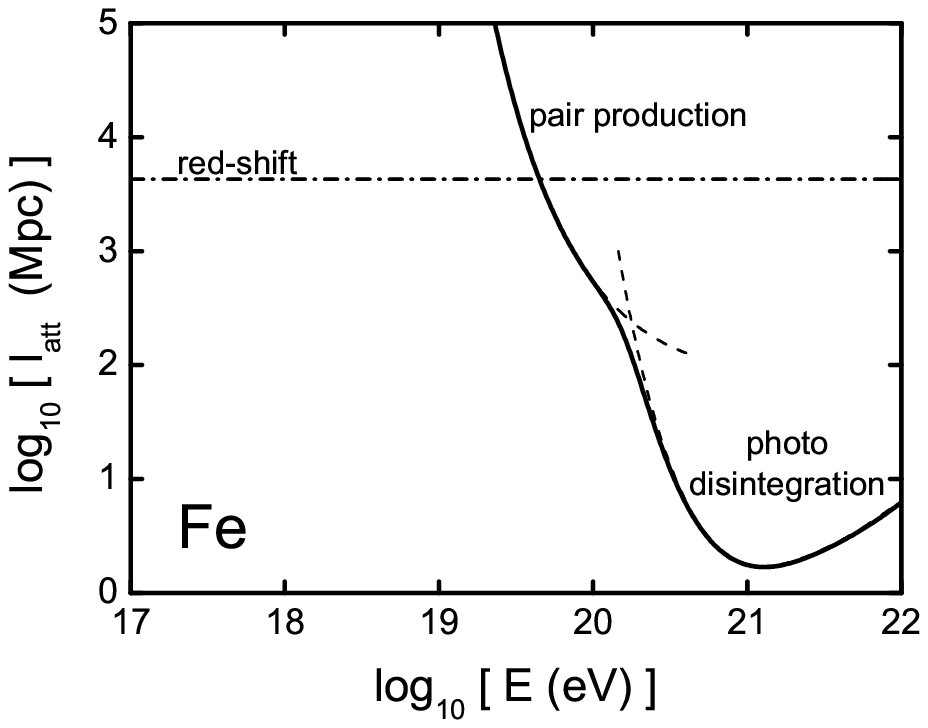}
&
\includegraphics[width=0.48\textwidth]{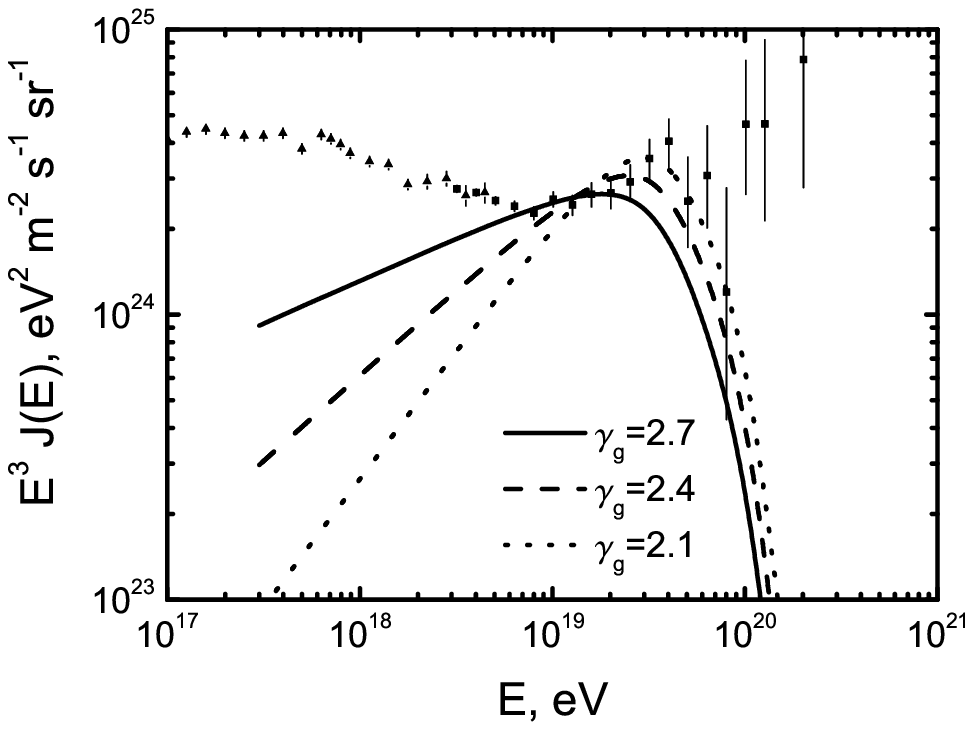}
\\
\end{tabular}
\caption{[Right Panel] Iron energy losses on the CMB background field.
[Left Panel] Spectra of Iron for different choices of the injection spectrum, 
$\gamma_g=2.7$ continuos line, $\gamma_g=2.4$ dashed line and $\gamma_g=2.1$ dotted line. The experimental points are the Akeno-AGASA data.}
\label{fig:nuclei}
\end{center}
\end{figure}

The description of the dip in terms of the pair production process fails if a substantial fraction of nuclei heavier than protons is included at injection \cite{BGG,dip,Allard}. In figure \ref{fig:nuclei} (left panel) we show the energy losses of iron nuclei on the CMB background. Apart from the adiabatic energy losses nuclei suffer also the pair production process and the photodisintegration process. Comparing the left panel of figure \ref{fig:protons} and \ref{fig:nuclei} it is evident how the pair production process works in the two different cases. In the case of protons pair production is efficient in an energy interval that covers almost one decade, while in the case of iron (but the same result holds for all nuclei \cite{nuclei}) the pair production process dominates energy losses only in a very short energy interval. Apart from the CMB background, nuclei propagation is also affected by the interaction with the IR/V/UV background, as discussed in \cite{Stecker,Allard}. The effect of this interaction is mainly related to the photodisintegration  process and does not affects our conclusions \cite{nuclei}.

The discussed difference in the energy losses of protons and nuclei is responsible for a completely different behavior of the expected fluxes. This result is shown in the right panel of figure \ref{fig:nuclei}, an iron dominated injection spectrum can reproduce the observations only at high energies and does not show the dip behavior. In a more quantitative way, already a fraction of 10 - 20\% of nuclei in the primary flux affects the best fit value $\gamma_g=2.7$ spoiling the good agreement with observations at the dip energies \cite{dip}.

\section{Transition from Galactic to Extragalactic Cosmic Rays}
\label{sec:transition}

As shown in the right panel of figure \ref{fig:protons}, at low enough energies the computed best fit spectrum becomes much lower than the observed all particle spectrum. This can be interpreted as an indication that a new component, of different origin, is contributing to the flux at these energies. We interpret this new component as due to the galactic cosmic rays. 

In the dip model the transition between galactic and extragalactic component takes place at energies below the critical energy\cite{dip} $E_c \approx 1\times 10^{18}$ eV. The critical energy $E_c$ is fully determined by the combination of the pair production losses and adiabatic losses due to the expansion of the universe (see figure \ref{fig:protons} left panel). These two channels of energy loss occur at the same rate at the energy\cite{AB} $E_{\rm eq}= 2.3\times 10^{18}$ eV. In a semi-quantitative way, the connection between $E_c$ and $E_{\rm eq}$ can be expressed as $E_c=E_{\rm eq}/(1+z_{\rm eff})^2$, where $z_{eff}$ is an effective redshift of the sources contributing to the flux of cosmic rays at energy $\sim E_c$. A simplified analytical estimate for $\gamma_g=2.6 - 2.8$ gives $1+z_{\rm eff} \approx 1.5$ and hence $E_c \approx 1\times 10^{18}$ eV. From the experimental point of view, the transition is expected to appear in the form of a second knee in the spectrum, at an energy $E_{\rm 2kn}$. Different experiments have found evidence of such second knee at energy $E_{\rm 2kn} \sim (0.4 - 0.8)\times 10^{18}$ eV.  A second knee is also expected in the case of the mixed composition model \cite{Allard} as a result of the superposition of a steep galactic spectrum and a flat extragalactic spectrum.

\begin{figure}[ht]
  \begin{center}
    \includegraphics[width=14.0cm]{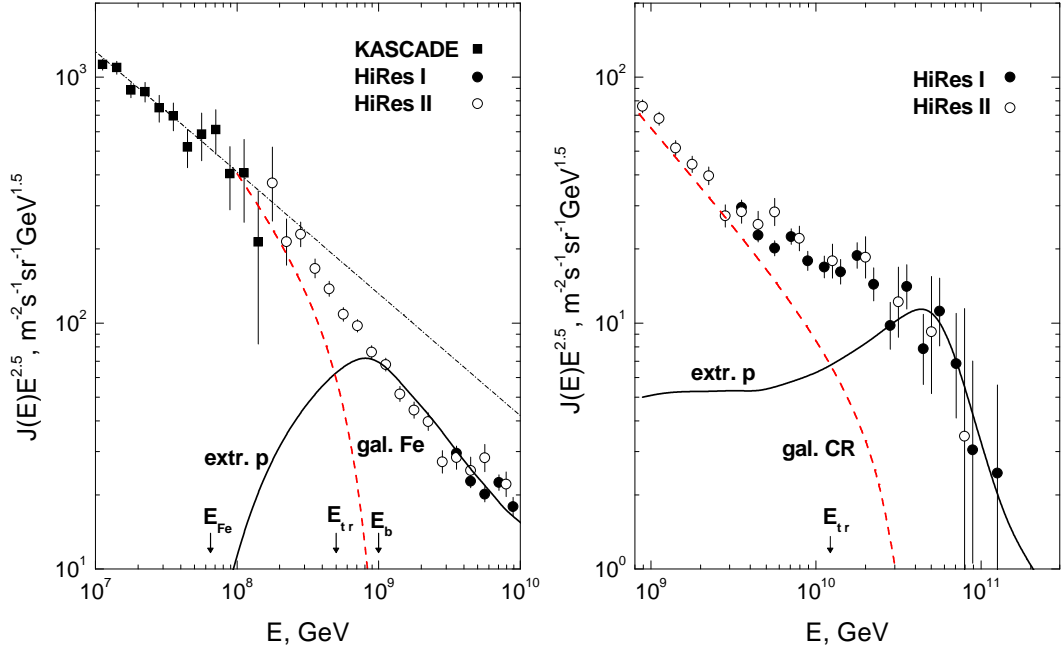}
\end{center}
  \caption{[Left Panel] Spectrum of extragalactic cosmic rays
  with injection spectrum $E^{-2.7}$, turbulent magnetic field $1$ nG
  with Bohm diffusion coefficient and closest source at 50 Mpc from
  the Earth. [Right Panel] Spectrum of extragalactic cosmic rays
  with injection spectrum $E^{-2}$, showing the transition at the
  ankle. In both cases the dashed line is obtained as a result of
  subtracting the extragalactic spectrum from the observed
  all-particle spectrum.}
  \label{fig:transition}
\end{figure}

This effect can be understood simply because in both transition models the extragalactic spectrum in the transition region is flatter than the spectrum of galactic cosmic rays. 

The scenario with a transition from galactic to extragalactic CR at the ankle is illustrated in the right panel of figure \ref{fig:transition}, where we plot the data from KASCADE, HiResI and HiRes II, the predicted spectrum from extragalactic sources with $\gamma_g=2.0$ (solid line) and the spectrum of galactic cosmic rays, calculated subtracting the predicted extragalactic flux from the all-particle observed spectrum.

For the case of transition through a second knee and a dip, the situation is depicted in the left panel of figure \ref{fig:protons}, at low enough energies, the spectrum of extragalactic protons reproduces the injection spectrum, which is always flatter than the spectrum of galactic 
cosmic rays $\propto E^{-3.1}$. This conclusion is strengthened by the presence of a magnetic field in the intergalactic medium, because of the fact that the propagation time from the nearest source may exceed the age of the universe, reflecting in the appearance of an exponential cutoff at low energy \cite{AB}. This case is illustrated in the left panel of figure \ref{fig:transition}, obtained for a turbulent magnetic field of $1$ nG and a Bohm diffusion coefficient.

One can clearly see from figure \ref{fig:transition} that the model of transition at the ankle requires a maximum energy of the galactic cosmic rays which exceeds $\sim 10^{19}$ eV. If the transition takes place at the second knee, then the maximum energy of galactic cosmic rays is predicted to be $\sim 10^{17}$ eV. In fact in both cases these fluxes are still appreciable at 
$\sim 4\times 10^{19}$ eV and $\sim 5\times 10^{17}$ eV respectively.

The most valuable insight in the problem of the transition comes from the KASCADE and Tibet \cite{dataGCR} experiments. The results on the chemical composition of cosmic rays from both experiments are rather strongly dependent on the Monte Carlo technique used for the reconstruction of the primary composition. The most solid conclusions seem to be those concerning the lighter nuclei, hydrogen and helium. 

\begin{figure}[ht]
\begin{center}
\begin{tabular}{ll}
\includegraphics[width=0.5\textwidth,height=0.4\textwidth]{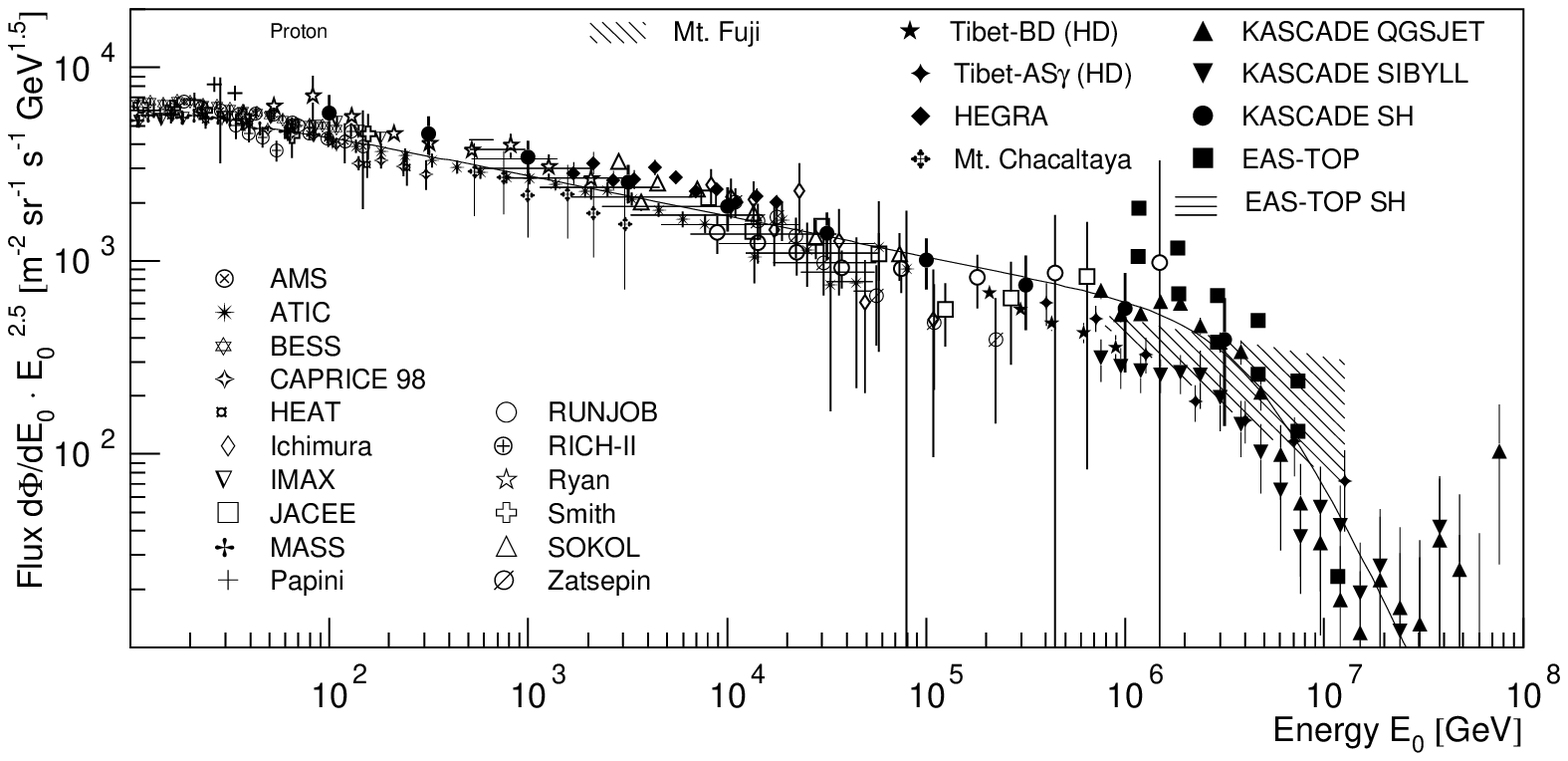}
&
\includegraphics[width=0.5\textwidth,height=0.4\textwidth]{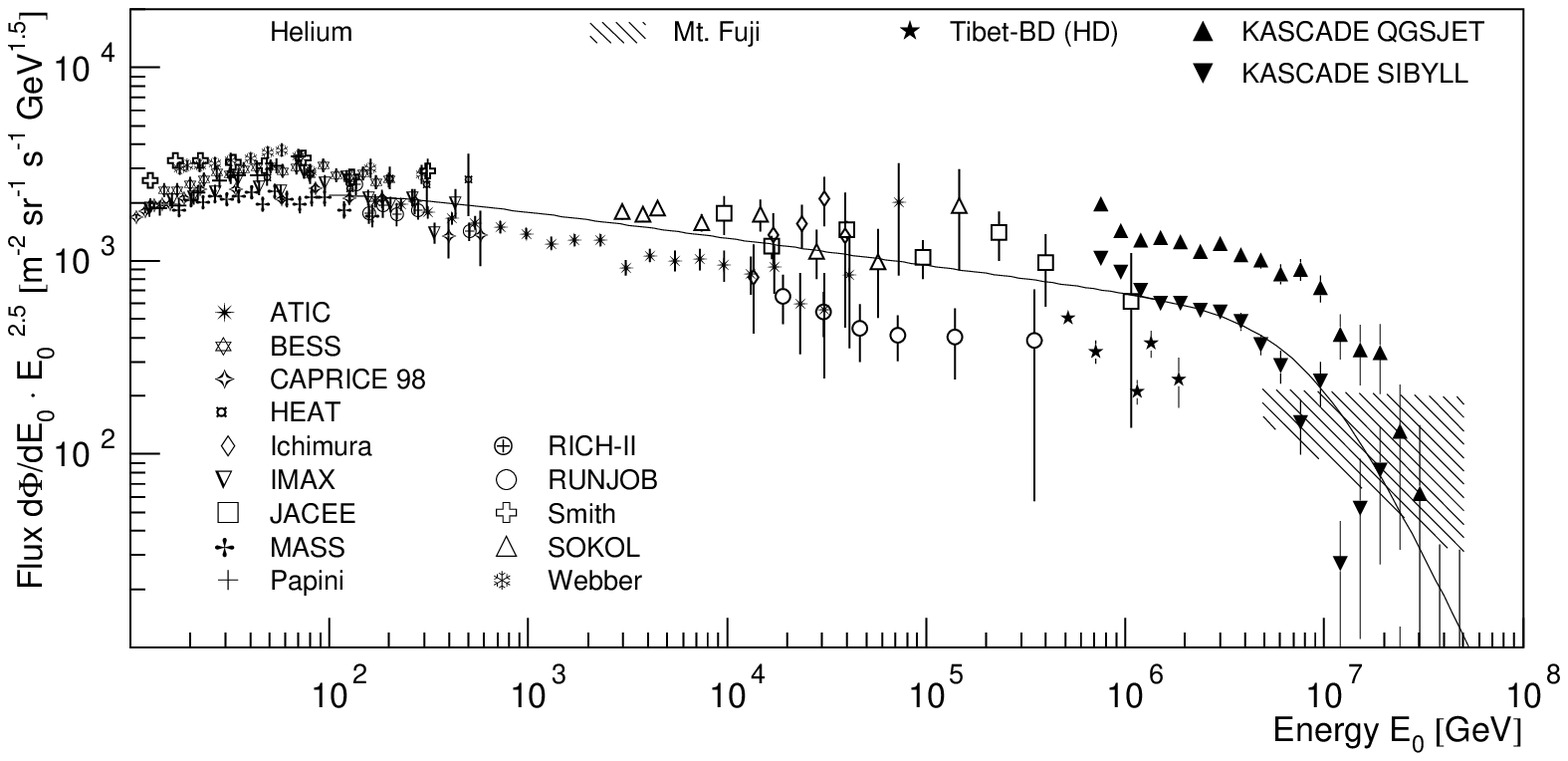}
\\
\end{tabular}
\caption{[Right Panel] Observed proton spectrum as function of the energy.
[Left Panel] Observed helium spectrum as function of the energy.}
\label{fig:kas}
\end{center}
\end{figure}

The error bars and the dependence of the results upon the choice of QJSJET \cite{QJSJET} or SYBILL \cite{SYBILL} for the simulation of the interactions of nuclei makes the measurements of more difficult interpretation when referred to heavier nuclei. The spectra of protons and helium measured by KASCADE and other experiments are shown in figure \ref{fig:kas} (from \cite{horandel}). A few comments are in order: 1) the spectrum of protons extends to at least $\sim 4\times 10^{16}$ eV, one order of magnitude larger than the
knee in the all-particle spectrum and roughly one order of magnitude higher than the knee in the proton component. 2) The slope of the spectrum of protons has a slope $\sim 2.7$ below the proton knee and $\sim 3.8$ above the proton knee. 3) Similar comments apply to helium
nuclei, where the knee appears to be at slightly higher energies.

At some point, the flux of protons should suffer a cutoff at the maximum energy of the accelerated protons. These energies, as discussed in \cite{blasirev}, exceed the prediction of quasi-linear theory applied to the case of supernova remnants, but would be consistent with more recent calculations \cite{Bell}. It is not clear what is the physical phenomenon that generates the knees in the single chemical components. If the mechanism is rigidity dependent, as suggested by the KASCADE data, one can expect a knee in the iron component at the energy $E\sim 6\times 10^{16}$, while the spectrum of iron would possibly end around $\sim 6\times 10^{17}$ eV.

In this picture, the knee in the all-particle spectrum is due to the superposition of the knees from different chemical components, and the galactic cosmic ray spectrum would disappear at roughly $\sim 6\times 10^{17}$ eV, although it would start to fade around $\sim 10^{17}$ eV. Although suggested by observations, this scenario is not unambiguosly proven by the KASCADE data or any other set of data at the present time. 

In the mixed composition scenario, as discussed above, it is required that the galactic component of the cosmic rays extends to energies that exceed $10^{19}$ eV, which appears to be in conflict with the scenario depicted above. This problem is often addressed by postulating that the actual ankle is at slightly lower energies, $3\times 10^{18}$ eV, or by assuming the existence of another galactic component, which appears at energies above the iron knee. In the case of the transition at the second knee, as seen in the left panel of figure \ref{fig:transition}, 
the maximum energy of the galactic component is perfectly compatible with the rigidity dependent extrapolation of the KASCADE data.

Finally, there is an additional issue, related to the measurement of the proton abundance by the Akeno detector \cite{dataUHECR}. The claim is that $\sim 10\%$ of the cosmic ray flux at $\sim 10^{17}$ eV is made of protons \cite{dataUHECR}. While in the mixed composition scenario this result would be very hard to interpret, it would find a natural explanation in the scenario 
described in section \ref{sec:thedip}. From the observational point of view, this point serves as a strong suggestion that the crucial discriminant between the mixed composition and the dip scenario is the different chemical composition expected in the transition region 
$10^{17}-10^{19}$ eV.

\section{Conclusions}
\label{sec:conclusions}

The dip is a feature in the spectrum of cosmic rays of extragalactic origin, that originates from Bethe-Heitler pair production of protons on the cosmic microwave background\cite{BGG,dip}. A dip-like feature is observed in the data of all current experiments. This feature is fitted in an excellent way by the predicted spectrum of extragalactic cosmic rays with injection spectrum with slope $2.7$, and would represent a precious indication of the extragalactic origin of cosmic rays down to energies of the order of $\sim 10^{18}$ eV. On the other hand, a similar feature is also generated in the context of the mixed composition scenario for the transition between galactic and extragalactic cosmic rays. In this scenario, the required
extragalactic injection spectrum is flatter than in the dip scenario and the required maximum energies of the galactic component are as high as $10^{19}$ eV. Both approaches to the problem have weak points, as recently discussed in \cite{Allard}, but none of them can at present serve to discriminate between the two. 

The mixed composition scenario appears more compatible with the predicted
injection spectrum as obtained from the theory of particle acceleration at shock fronts. It can also accomodate more easily the presence of heavy nuclei in the chemical composition of extragalactic cosmic rays. However, if the transition between a galactic and an extragalactic origin of cosmic rays occurs at $E>10^{18}$ eV, then the maximum energy of galactic cosmic rays appears to be in contradiction with a smooth extrapolation of the KASCADE data, as well as with the standard theory of the origin of galactic cosmic rays. 

On the other hand, the dip scenario apparently implies a large energetics for the sources of extragalactic cosmic rays as a result of the steep injection spectra required to fit the data \cite{dip}. However, as discussed in \cite{dip}, there are numerous ways to solve these problems without invoking exotic or unnatural solutions. More serious is the problem related to the fact that the dip exists only if the fraction of nuclei at the sources, and helium in particular, is kept rather low. Only accurate measurements of the chemical composition in the energy region between $10^{17}$ eV and $5\sim 10^{18}$ eV can provide a clue to the applicability of the two alternative scenarios.

\section*{Acknowledgments}
I am grateful to V. Berezinsky, P. Blasi, S. Grigorieva, A. Gazizov and B. Hnatyk with whom the present work was developed.

\end{document}